\documentclass[conference]{IEEEtran}
\IEEEoverridecommandlockouts

\usepackage{cite}
\usepackage{amsmath,amssymb,amsfonts}
\usepackage{algorithmic}
\usepackage{graphicx}
\usepackage{textcomp}
\usepackage{xcolor}
\usepackage{comment}

\usepackage[pscoord]{eso-pic}
\newcommand{\placetextbox}[3]{
  \setbox0=\hbox{#3}
  \AddToShipoutPictureFG*{
    \put(\LenToUnit{#1\paperwidth},\LenToUnit{#2\paperheight})%
    {\vtop{{\null}\makebox[0pt][c]{#3}}}}
  }
\placetextbox{.2}{0.055}{978-3-903176-78-2 \textcopyright\ 2026 IFIP}

\def\BibTeX{{\rm B\kern-.05em{\sc i\kern-.025em b}\kern-.08em
    T\kern-.1667em\lower.7ex\hbox{E}\kern-.125emX}}
\begin{document}

\title{Selective Placement of Hollow-Core Fibers for \\ QKD and Classical Communication Coexistence}

\author{Giovanni Simone Sticca$^*$, Alessandro Gagliano, Memedhe Ibrahimi, \\ Alberto Gatto, Francesco Musumeci, and Massimo Tornatore}
\author{\IEEEauthorblockN{Giovanni Simone Sticca$^*$, Alessandro Gagliano, Memedhe Ibrahimi, \\ Alberto Gatto, Francesco Musumeci, and Massimo Tornatore}
\IEEEauthorblockA{\textit{Politecnico di Milano, Milan, Italy};  $^*$Corresponding author: giovannisimone.sticca@polimi.it}
}

\maketitle

\begin{abstract}
We investigate the benefits of partially upgrading optical networks with hollow-core fibers for QKD-classical communication coexistence. Results show that upgrading 40\% of links in a metro topology can reduce the number of quantum modules by up to 49\%. 
\end{abstract}


\section{Introduction}
\vspace{-2pt}
Quantum key distribution (QKD) is a cornerstone technology for future secure communication networks, offering information-theoretic security against classical and quantum adversaries. However, its large-scale deployment remains challenging due to the high cost of dedicated infrastructures~\cite{cai2021simultaneous}. A more practical approach is the integration of QKD and classical communications into the same optical network, where quantum and classical signals coexist over the same fiber. 
Yet, in standard single-mode fibers (SSMFs), the coexistence of weak quantum signals with classical channels in the C-band is severely limited by non-linear noise generated by classical signals, resulting in reduced secret key rates (SKRs) and shorter feasible transmission distances. 



A novel and promising technology that ensures low transmission loss and low nonlinearities is Hollow-Core Fiber (HCF). While still in its infancy, HCF has shown significant benefits for Data Center Interconnection (DCI), thanks to its reduced propagation latency compared to SSMF. HCF is also gaining momentum in quantum communications with British Telecom conducting the world’s first field trial of quantum-secure communications over HCF in 2021~\cite{bt2021trial}, and the latest experiments over HCFs have confirmed simultaneous quantum–classical transmission with greatly reduced Raman noise~\cite{dou2025hcf}. 
In particular, HCF exhibits a much lower Raman scattering efficiency than SSMF~\cite{honz2023first}, enabling the transmission of quantum channels in the C-band without being disrupted by co-propagating classical signals. In addition, the low loss of HCF enables longer transmission distances and higher SKRs compared to SSMF.
While the challenge to deploy HCF in today's networks remains its high cost, its deployment in QKD–classical communication coexistence scenarios might be justified, since it enables a significant reduction in the number of costly quantum modules.

\begin{figure}[t]
\centering
\includegraphics[width=0.8\linewidth]{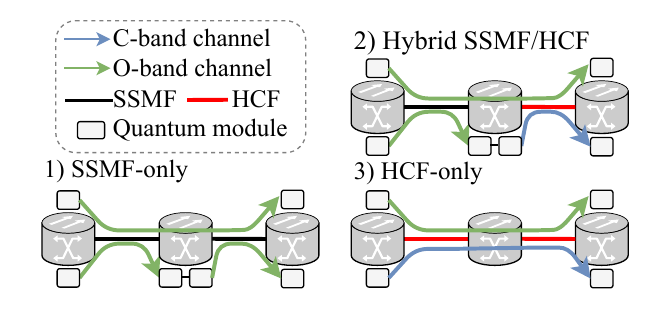}
\vspace{-15pt}
\caption{Quantum channel allocation example in a \textit{hybrid SSMF/HCF} network.}
\vspace{-18pt}
\label{fig:example}
\end{figure}

In this work, we investigate the role of HCF in enabling efficient QKD–classical communication coexistence from a network-wide perspective. 
Since HCF deployment is currently limited by high cost and limited availability, we consider three possible configurations for fiber deployment (namely \textit{SSMF-only}, \textit{hybrid SSMF/HCF}, and \textit{HCF-only}), where different combinations of band assignment can be used to achieve co-existence of classical and QKD channels, as shown in Fig.~\ref{fig:example}. Specifically, in the \textit{SSMF-only} configuration, all the links are SSMF, and quantum channels are operated only in the O-band, to avoid Raman noise from classical signals operating in C-band; quantum channels can be either end-to-end or realized using intermediate relays (i.e., using two back-to-back quantum modules, where the channel is received and then re-transmitted). In the \textit{hybrid SSMF/HCF} configuration, quantum channels can operate only in the O-band over SSMF spans, and both in the C- or O-band over HCF segments, thanks to the strongly reduced Raman scattering in HCF. Finally, in the \textit{HCF-only} configuration, both C- and O-band quantum transmission are feasible. To evaluate the benefits of the partial HCF deployment that characterizes the \textit{hybrid SSMF/HCF} configuration, we devise an optimization framework that allows to quantify how selective HCF placement can minimize the number of deployed quantum modules required to achieve a given SKR between network nodes. Numerical results on a realistic metro-scale topology show that even partial link upgrades to HCF, e.g., 30–40\%, are sufficient to achieve most of the benefits, reducing the number of required quantum modules by up to 49\% compared to an \textit{SSMF-only} network. 

\section{Secret Key Rate Evaluation}
\vspace{-3pt}
As a first step to develop our framework, we introduce the modeling of the physical-layer performance of quantum channels. We consider the discrete-variable BB84 protocol with decoy state method and finite-size effect based on~\cite{yin2020tight}. The achievable SKR depends on both the physical transmission properties of the fiber spans and the impairments introduced by co-propagating classical channels.  

Quantum and classical signals are multiplexed on a common WDM grid with 100 GHz channel spacing. In the C-band, we assume a bandwidth of $4$~THz (192--196~THz), within which a $1$~THz block centered at $194$~THz is reserved for quantum channels. This block is isolated from neighboring classical channels by $300$~GHz guard bands on each side, while the remaining portion of the C-band is used for classical traffic. 
Quantum channels can also be allocated in the O-band, spanning $4$~THz (232--236~THz). At each intermediate node or inline amplifier, quantum and classical signals are separated by multiplexing/demultiplexing components, each introducing $1$~dB insertion loss. The physical channel properties accounted for in the SKR calculation are fiber attenuation, spontaneous Raman scattering (SpRS), and four-wave mixing (FWM). Linear crosstalk is neglected, since the guard bands provide sufficient isolation. Both SpRS and FWM are evaluated on a per-span basis as in \cite{cai2021simultaneous,vorontsova2023theoretical}, considering that optical amplifiers restore classical channels' power to $0$~dBm at each span. 

The SKR computation distinguishes between SSMF and HCF. In particular, for SSMF, Raman efficiency and attenuation are considered as in~\cite{gagliano2024quantum}, whereas for HCF, we consider a reduced Raman efficiency (scaled by $35$~dB as in~\cite{honz2023first}) and the attenuation profile as in \cite{polettiOFC24}. The final SKR is then obtained by combining transmission efficiency, noise contributions, and finite-key effects, following the methodology described in~\cite{gagliano2024discrete}.

\begin{figure}[t]
\centering
\includegraphics[width=0.8\linewidth]{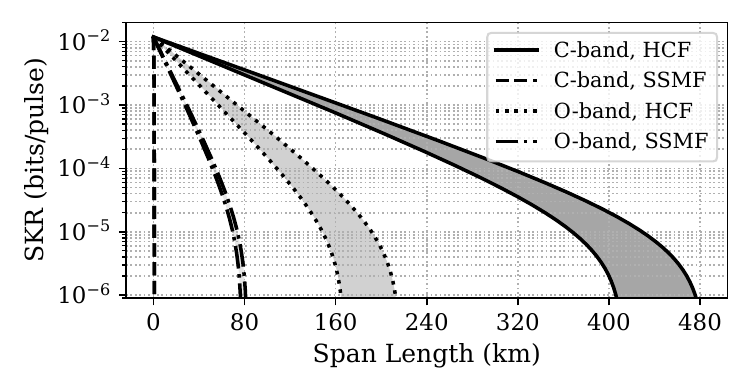}
\vspace{-15pt}
\caption{Secret key rate vs. span length.}
\vspace{-20pt}
\label{fig:skr}
\end{figure}

Figure~\ref{fig:skr} shows the SKR as a function of span length for quantum channels transmitted in either the C- or O-band, under the assumption of full classical channel loading, for both SSMF and HCF. For each case, the shaded region represents the range of SKR values obtained over all quantum channels in the considered band, bounded by the best- and worst-performing channels. The results show that coexistence in the C-band over SSMF is not feasible due to strong Raman noise from the classical signals, while the O-band supports metro-scale transmission (around 80 km). In contrast, HCF significantly extends the reach by about $5.6\times$, with C-band coexistence proving more effective than O-band thanks to the lower transmission loss in the C-band.
Although this analysis confirms the physical-layer benefits of HCF, a network-wide study is required to quantify the overall cost savings in terms of the number of quantum modules. Moreover, given the current high cost and limited availability of HCF due to the lack of mass production, it is important to assess whether a \emph{partial placement} strategy, in which HCF is selectively introduced on specific links, can deliver substantial benefits.

\section{Selective HCF Placement for QKD-Classical Communication Coexistence}
\vspace{-5pt}
\noindent \textbf{Problem statement}. 
Given a network topology represented as a graph $G(N,E)$, where $N$ is the set of nodes and $E$ the set of fiber links, and a set of quantum secret-key demands $T = \{t = (s_t, d_t, k_t)\}$, where each demand $t$ is defined by a source node $s_t \in N$, a destination node $d_t \in N$, and a requested secret key rate $k_t$, we decide the subset of links to be upgraded to HCF (replacing the existing SSMF). The objective is twofold: (\emph{i}) the primary objective is to maximize the fraction of served secret-key demands, thereby minimizing the unserved key rate, and (\emph{ii}) the secondary objective is to minimize the number of quantum modules deployed. 

\noindent \textbf{Genetic Algorithm (GA) for selective HCF placement}. 
We adopt a GA that 
begins with a randomly generated population of individuals, each representing a candidate solution. An individual is encoded as a binary vector of length $|E|$, where a value of $1$ indicates that the corresponding link is upgraded to HCF. A budget constraint restricts the number of link upgrades to at most $\lfloor b \cdot |E| \rfloor$, where $b \in [0,1]$ is the HCF budget fraction. 

Individuals are evaluated using a \emph{fitness function}, and the top-performing 40\% are selected for reproduction. The population is updated through standard GA operators: crossover (to recombine solutions), mutation (to introduce local variations), and random injection (to maintain population diversity). This process is repeated until the stopping criterion is satisfied, i.e., no improvement over a fixed number of generations.


The \emph{fitness function} quantifies the quality of each individual by applying the corresponding HCF placement to the graph $G$ and provisioning all demands $t \in T$. 
Each demand $t$ is routed along its shortest path $P_t$, where relays can be placed at intermediate nodes. Given that $P_t$ includes $N_{P_t}$ nodes, up to $N_{P_t}-2$ relays can be deployed. Relays are positioned so that the resulting $(r+1)$ path segments have approximately equal lengths, thereby balancing transmission impairments.
For each relay count $r \in \{0,1,\dots,N_{P_t}-2\}$, the algorithm evaluates the corresponding cost ${cost}_t^r$ and served key rate $k_{s,t}^r$. In particular, the presence of $r$ relays divides the path $P_t$ into $r+1$ contiguous subpaths, denoted as $\{S_{1,t}, S_{2,t}, \dots, S_{r+1,t}\}$. Since each subpath $S_{i,t}$ requires one transmitter and one receiver, channels can be allocated independently. For every $S_{i,t}$, the algorithm searches for one available channel in both the C-band and O-band using a first-fit strategy. 

\begin{figure*}[t]
\centering
\includegraphics[width=\textwidth]{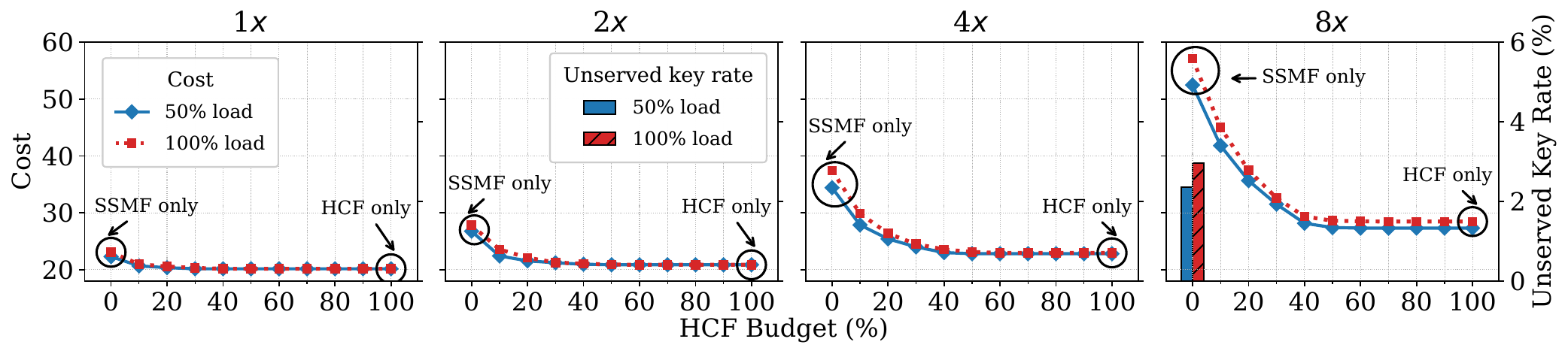}
\vspace{-25pt}
\caption{Average cost (number of quantum modules) and unserved key rate as a function of the HCF budget, for different link length factors ($1\times$, $2\times$, $4\times$, $8\times$) under \emph{half-load} and \emph{full-load} of classical channels.}
\vspace{-20pt}
\label{fig:tokyo-results}
\end{figure*}

Then, the achievable key rates in $S_{i,t}$ for the two bands, $k_c^{S_{i,t}}$ and $k_o^{S_{i,t}}$, are computed.
The band yielding the higher key rate is selected, the corresponding channel is occupied (increasing the channel counter $c_t^r$ by one), and the assigned key rate for that subpath is defined as $k^{S_{i,t}} = \max(k_c^{S_{i,t}}, k_o^{S_{i,t}})$. After all subpaths have been processed, the end-to-end achievable key rate is determined as the minimum among them, $k = \min_{i=1,\dots,r+1} k^{S_{i,t}}$, and the cumulative served rate is updated as $k_{s,t}^r \leftarrow k_{s,t}^r + k$. This process continues until either the demand is satisfied ($k_{s,t}^r \geq k_t$) or no further channels can be provisioned (due to insufficient available spectrum or exceeded maximum reach). The final cost for configuration $r$ is then computed as $cost_t^r = 2 \cdot c_t^r$, which accounts for two quantum modules for each channel. 
After evaluating all relay configurations, the algorithm selects the number of relays $r^\star$ that minimizes the unserved portion of demand $t$ ($u_t$). 

\begin{figure}[ht]
\centering
\includegraphics[width=\linewidth]{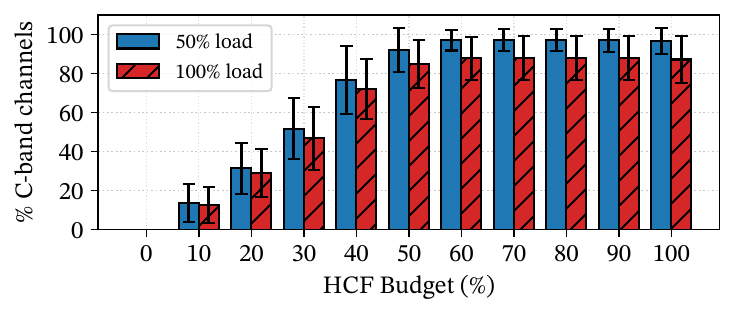}
\vspace{-25pt}
\caption{Percentage of C-band quantum channels vs. HCF budget, averaged across link-length multipliers (1×–8×).}
\vspace{-20pt}
\label{fig:C_band_perc}
\end{figure}

If multiple configurations achieve zero unserved traffic, the one with the lowest cost is chosen. Consequently, the cost and the unserved traffic for demand $t$ are $cost_t = cost_t^{r^\star}$ and $u_t = u_t^{r^\star}$, respectively.
Finally, the fitness of an HCF placement is the sum of the contributions across all demands: $f = \alpha \cdot \sum_{t \in T} u_t + \beta \cdot \sum_{t \in T} cost_t$. The weights $\alpha, \beta \geq 0$ are set to prioritize the minimization of unserved traffic,
i.e., $\alpha \gg \beta$.

\section{Numerical Results}
\vspace{-4pt}
We consider the Tokyo metro optical network~\cite{tachibana2023man}, consisting of $23$ nodes and $43$ links, with an average link length of $5.5$~km. A set of $10$ quantum secret-key demands is generated, each with a random source--destination pair and a secret key rate request uniformly distributed in $(2,20)$~kbps. For each demand set, we determine the optimal placement of HCF upgrades under different budget constraints, ranging from $0\%$ (all links deployed as SSMF, considered the baseline in our study) to $100\%$ (all links upgraded to HCF, representing the lower bound). To assess the impact of transmission distance, all link lengths are scaled by factors of $1\times$, $2\times$, $4\times$, and $8\times$. Two levels of classical channel load in the C-band are considered: \emph{half-load} ($50\%$) and \emph{full-load} ($100\%$). Each experiment is repeated 50 times with independent demand realizations. 
Figure~\ref{fig:tokyo-results} shows the number of quantum modules (cost) and the percentage of unserved key rate. 

In general, increasing the percentage of HCF links consistently reduces the cost, with savings becoming more significant as link lengths increase. In all cases, a partial HCF upgrade, in the range of $30$--$40\%$ of the links, is sufficient to reach the lower bound in terms of cost.
At $1\times$ length, all demands are served even without HCF upgrades. Increasing the HCF budget reduces the cost by up to $10\%$ under \emph{half-load} and $13\%$ under \emph{full-load}, with most of the savings already achieved at $30\%$ HCF budget. 
The cost reduction stems from the lower Raman efficiency of HCF, which mitigates interference between classical and quantum channels and allows the use of the C-band. 
As link lengths increase, cost savings become more significant. For example, at $2\times$ length, the cost reduction reaches up to $25\%$, while, at $4\times$ length, it grows to about $38\%$, compared to the \textit{SSMF-only} scenario, i.e., 0\% HCF budget. In both cases, most of the benefits are already captured with HCF budgets around $30$--$40\%$.
At $8\times$ length, in the \textit{SSMF-only} scenario, we cannot ensure that all quantum traffic requests are served, i.e., we observe between 2-3\% unserved key rate at \emph{half-load} and \emph{full-load}.
Introducing HCF in 10\% of the links, all quantum traffic requests are served, and we achieve cost savings of over 20\% at both \emph{half-load} and \emph{full-load}.
Increasing the HCF budget to $40\%$ further enhances savings, reaching $46\%$ at \emph{half-load} and $49\%$ at \emph{full-load}. 
Additional upgrades (over 40\%) bring only marginal improvements.

Let us now consider the percentage of quantum channels allocated in the C-band, as shown in Fig.~\ref{fig:C_band_perc}. As a partial upgrade with HCF provides significant benefits in terms of cost, a byproduct of introducing HCF is that it allows increasing the percentage of quantum channels in the C-band. For instance, at a 40\% HCF budget, the percentage of C-band quantum channels ranges between $72\%$ and $77\%$. Nevertheless, a portion of channels remains in the O-band, since in hybrid paths traversing both SSMF and HCF segments, operating in the O-band avoids the SpRS noise generated in SSMF sections. These results highlight that, in \textit{hybrid SSMF/HCF} networks, a balanced allocation of quantum channels across both the C- and O-bands is essential to minimize overall cost.




\vspace{-5pt}
\bibliographystyle{IEEEtran}
\bibliography{bibliography.bib}

@ARTICLE{cai2021simultaneous,
  author={Cai, Chun and others},
  journal={IEEE TCOM},
  title={Simultaneous Long-Distance Transmission of Discrete-Variable Quantum...},
  year={2021},
  volume={69},
  number={5},
  pages={3222-3234},
  doi={10.1109/TCOMM.2021.3056528}
}

@inproceedings{dou2025hcf,
  author={Dou, Tianqi and others},
  year={2025},
  booktitle={ECOC},
  month={09},
  pages={1-4},
  title={{Experimental Demonstration of $47 \times 800 \text{Gbps}$ Classical Communication...}},
  doi={10.1109/ECOC66593.2025.11263383}
}

@article{yin2020tight,
  title={Tight Security Bounds for Decoy-State Quantum Key Distribution},
  volume={10},
  ISSN={2045-2322},
  DOI={10.1038/s41598-020-71107-6},
  number={1},
  journal={Scientific Reports},
  publisher={Springer Science and Business Media LLC},
  author={Yin, Hua-Lei and others},
  year={2020},
  month={aug}
}

@ARTICLE{gagliano2024quantum,
  author={Gagliano, Alessandro and others},
  journal={IEEE TCOM},
  title={Quantum Key Distribution Spectral Allocation and Performance...},
  year={2025},
  volume={73},
  number={1},
  pages={510-523},
  doi={10.1109/TCOMM.2024.3439447}
}

@article{honz2023first,
  author={Honz, Florian and others},
  year={2023},
  month={06},
  pages={1-7},
  title={First Demonstration of 25$\lambda$ × 10 Gb/s C+L Band Classical / DV-QKD Co-Existence...},
  volume={PP},
  journal={JLT},
  doi={10.1109/JLT.2023.3256352}
}

@INPROCEEDINGS{polettiOFC24,
  author={Chen, Y. and others},
  booktitle={OFC},
  title={Hollow Core DNANF Optical Fiber with $<$0.11 dB/km Loss},
  year={2024},
  pages={1-3}
}

@ARTICLE{gagliano2024discrete,
  author={Gagliano, Alessandro and others},
  journal={JOCN},
  title={Discrete-Variable Quantum Key Distribution Services ...},
  year={2025},
  volume={17},
  number={1},
  pages={A96-A102},
  doi={10.1364/JOCN.534366}
}

@ARTICLE{tachibana2023man,
  author={Nakamura, Koga and others},
  journal={IEEE Access},
  title={Metropolitan Area Network Model Design Algorithm...},
  year={2025},
  volume={13},
  pages={82503-82513},
  doi={10.1109/ACCESS.2025.3568510}
}

@misc{bt2021trial,
  author={{BT press release}},
  title={},
  year={},
  note={https://newsroom.bt.com/bt-conducts-worlds-first-trial-of-quantum-secure-communications-over-hollow-core-fibre-cable/}
}

@article{vorontsova2023theoretical,
  author={Irina Vorontsova and others},
  journal={J. Opt. Soc. Am. B},
  number={1},
  pages={63--71},
  publisher={Optica Publishing Group},
  title={Theoretical Analysis of Quantum Key Distribution Systems...},
  volume={40},
  month={Jan},
  year={2023},
  doi={10.1364/JOSAB.469933}
}
\end{document}